\documentstyle [12pt] {article}

\catcode`@=12
\newcommand{\be}{\begin{equation}}
\newcommand{\ee}{\end{equation}}
\newcommand{\ber}{\begin{eqnarray}}
\newcommand{\eer}{\end{eqnarray}}
\newcommand{\bers}{\begin{eqnarray*}}
\newcommand{\eers}{\end{eqnarray*}}
\begin{document}
\vspace{0.5in}
\oddsidemargin -.375in
\def\ee{\end{equation}}
\thispagestyle{empty}
\begin{flushright} ISU-NP-99-08\\
IITAP-99-010\\
September 1999\
\end{flushright}
\vspace {.5in}
\begin{center}
{\Large\bf  Semileptonic and non-leptonic $B_c$ decays \\}
\vspace{.2in}
{\bf A. Abd El-Hady ${}^{a,d}$, 
J. H. Mu\~noz ${}^{b,d}$ and J.~P.~Vary ${}^{c,d}$\\}
\vspace{.2in}
${}^{a)}$  
{\it
Physics Department, Zagazig University, Zagazig, Egypt}\\
${}^{b)}$  
{\it
Departamento de F\'{\i}sica, Universidad del Tolima, A. A. 546, Ibagu\'e, Colombia} \\ 
${}^{c)}$  
{\it
Department of Physics and Astronomy, Iowa State University , Ames, Iowa 50011, USA}\\
${}^{d)}$  
{\it
International Institute of Theoretical and Applied Physics,\\
Iowa State University, Ames, Iowa 50011, USA}\\

\vskip .5in
\end{center}

\vskip .1in \begin{abstract} 

We make predictions for the exclusive semileptonic and the non-leptonic decay
widths of the $B_c$ meson. We evaluate the $B_c$ semileptonic form
factors for different decay channels in a relativistic model, and use
factorization to obtain the non-leptonic decay widths.

\end{abstract}
\vskip .25in
\newpage

%
%

The recent discovery of the $B_c$ meson by the CDF Collaboration
\cite{CDF} attracted a great deal of attention. The $B_c$ meson is very
interesting because it carries non-vanishing flavor quantum numbers, and
lies below the threshold of the $BD$ decay. Therefore, it can only decay
through weak interactions which makes this doubly heavy meson useful for
studying the weak decays of heavy flavors. The $B_c$ production
mechanisms, spectroscopy, and decays have been analyzed using different
approaches ( see Ref. \cite{review} for a review ).

In a previous paper \cite{pp} we have used a relativistic model
\cite{sommerer2} based on the Bethe-Salpeter
Equation (BSE) to evaluate the spectrum of the $B_c$ meson. No free
parameters were used to fit the $B_c$ spectrum. Instead, all the model
parameters had been fixed in previous investigations of other meson
spectra. We also evaluated the decay constant of the $B_c$ meson, the
inclusive decay widths of the c-quark and the $\bar {b}$-quark together
with the annihilation width. Our results agree very well with the CDF
results of the $B_c$ mass and lifetime. We have presented these results
with two covariant reductions of the BSE and observed little dependence
on the choice of the reduction especially in the heavy flavor sector.

In this paper we evaluate the exclusive semileptonic $B_c \rightarrow
P(V) e \nu$ and two-body non-leptonic $B_c \rightarrow PP, \ PV, \ VV$
decay widths, where P (V) denotes a pseudoscalar (vector) meson. We use
our model to calculate the semileptonic form factors for different decay
channels. We then use factorization to obtain the non-leptonic decay
widths. We will utilize primarily a single reduction since this
investigation uses BSE results from the heavy flavor sector.

The BSE provides an appealing starting point to describe hadrons as
relativistic bound states of quarks. The BSE
for a bound state may be written in momentum space in the form
\begin{eqnarray}
G^{-1}(P,p)\psi(P,p)=\int\frac{1}{(2\pi)^{4}}V(P,p-p')\psi(P,p')d^4p',
\end{eqnarray}  
where $P$ is the four-momentum of the bound state, $p$ is the relative
four-momentum of the constituents . The BSE has three elements, the
interaction kernel ($V$) and the propagator ($G$) which we provide as
input, and the amplitude ($\psi$) obtained by solving the equation. We
also solve for the energy, which is contained in the propagator.

Different approaches have been developed to make the four dimensional
BSE more tractable and physically appealing. These include the
Instantaneous Approximation (IA) and Quasi-Potential Equations (QPE)
\cite{itzykson}. In the IA, the interaction kernel is
taken to be independent of the relative energy. In QPE, the two particle
propagator is modified in a way which keeps covariance and reduces the
4-dimensional BSE to a 3-dimensional equation. Of course, there is
considerable freedom in carrying out this reduction.

Earlier, we have used two reductions of the QPE to study the meson spectrum
\cite{sommerer2}. These reductions correspond to different choices of the
two particle propagator used to reduce the problem into three
dimensions. We refer to these reductions as A and B.  Reduction A
corresponds to a spinor form of the Thompson equation \cite{Thompson}
and reduction B corresponds to a new QPE introduced in
Ref. \cite{sommerer}. These two covariant reductions are chosen because they are
shown to give good fits to the meson spectrum.
In both reductions, we assume the interaction kernel to consist of
a one gluon exchange interaction, $V_{OGE}$, in the ladder
approximation, and a phenomenological, long range scalar confinement
potential, $V_{CON}$ given in the form
\begin{eqnarray}
V_{OGE}+V_{CON} & = & -{4\over
3}\alpha_s{\gamma_\mu\otimes\gamma_\mu\over {(p-p')^2}}
+\sigma{\rm\lim_{\mu\to 0}}{\partial^2\over\partial\mu^2} {{\bf
1}\otimes{\bf 1}\over-(p-p')^2+\mu^2}. \ 
\end{eqnarray}
Here, $\alpha_s$ is the strong coupling, which is weighted by the meson
color factor of ${4\over 3}$, and the string tension $\sigma$ is the
strength of the confining part of the interaction.  We adopt a scalar
Lorentz structure $V_{CON}$ as discussed in \cite {sommerer2}. In
our formulation of BSE there are a total of seven parameters : four
masses, $m_{u}$=$m_{d}$, $m_s$, $m_{c}$, $m_{b}$; the string tension
$\sigma$, and two other parameters used to govern the running of the
strong coupling constant. We varied these parameters to get the best fit
for a list of known mesons as described in \cite{sommerer2}.

In our subsequent work \cite{pp} on the $B_c$ meson, we evaluated the
$B_c$ spectrum without changing the parameter values mentioned above (
see Table \ref{tp810} below ) and compared our results with those of
Eichten and Quigg \cite{EQ} and Gershtein {\it at al.} \cite{GER} using
both the Martin potential and Buchmuller-Tye (BT) potential. The first
row of Table \ref{tp810} should be compared with the experimental result
\cite{CDF} of 6.40 $\pm$ 0.39 (stat.) $\pm$ 0.13 (syst.)  GeV/$c^2$. We
have also evaluated the inclusive $c$-quark and $\bar{b}$-quark decay
lifetimes \cite{pp} and obtained a $B_c$ lifetime of 0.46-0.47 ps in
good agreement with the experimental $B_c$ lifetime of
$0.46^{+0.18}_{-0.16}$ (stat.) $\pm$ 0.03 (syst.) ps \cite{CDF}.

\begin{table}[h!tb]
\caption{Spectrum of $B_c$ mesons in different channels (GeV/$c^2)$.}
\begin{center}
\begin{tabular}{|c|c|c|c|c|c|c|}
\hline\hline
State & Our work   & Our work   & Eichten and Quigg &Gershtein {\it et al.}\cite{GER}&
Gershtein {\it et al.}\cite{GER}\\
      & Reduction A & Reduction B & Ref. \cite{EQ} & Martin potential               &
BT potential                     \\
\hline\hline
$1^1S_0$& 6.356 & 6.380 & 6.264 & 6.253 & 6.246 \\
$1^3S_1$& 6.397 & 6.415 & 6.337 & 6.317 & 6.337 \\
$1^3P_0$& 6.673 & 6.692 & 6.700 & 6.683 & 6.700 \\
$1^3P_2$& 6.751 & 6.773 & 6.747 & 6.743 & 6.747 \\
$1^1P_1$& 6.752 & 6.777 &       & 6.729 & 6.736 \\
$2^1S_0$& 6.888 & 6.874 & 6.856 & 6.867 & 6.856 \\
$2^3S_1$& 6.910 & 6.891 & 6.899 & 6.902 & 6.899 \\
$1^3D_1$& 6.984 & 6.955 & 7.012 & 7.008 & 7.012 \\
\hline\hline
\end{tabular}
\label{tp810}
\end{center} 
\end{table}

We now turn our attention to exclusive decays. The $B_c$ exclusive
semileptonic and non-leptonic decays have been discussed in the
literature \cite{GER, ex1, cc}. The effective Hamiltonian for the semileptonic
decays has the standard current-current form, and is given by
\begin{eqnarray}
H_{W} & = & \frac{G_F}{\sqrt{2}}V _{ Q q } {\bar q}\gamma_{\mu}(1-\gamma_{5}) Q
{\bar{\nu}}\gamma^{\mu}(1-\gamma_{5})l .\ 
\end{eqnarray}
The leptonic current is completely known and the matrix element of the
vector ($V_{\mu}$) and the axial vector ($ A_{\mu}$) hadronic currents
between the meson states are represented in terms of form factors which
are defined by (considering the channel $B_c \rightarrow B_s
(B_s^{*})$)
\begin{eqnarray}
\langle B_s(P^{'})| V_\mu |B_c(P) \rangle\ &=&
f_+(P+P^{'})_\mu+f_-(P-P^{'})_\mu , \nonumber\\ 
\langle B_s^*(P^{'},\varepsilon)| V_\mu |B_c(P) \rangle\ &=&
ig\epsilon_{\mu\nu\alpha\beta}\varepsilon^{*\nu}(P+ P^{'}) ^\alpha
(P-P^{'})^\beta , \nonumber\\ 
\langle B_s^*(P^{'},\varepsilon)| A_\mu |
B_c(P) \rangle\ &=& f\varepsilon^*_\mu+(\varepsilon^* . P)[a_+
(P+P^{'})_\mu+a_-(P-P^{'})_\mu] .\ 
\label{slbc}
\end{eqnarray}
$f_{+}, f_{-}, g, f, a_+$ and $a_-$ are Lorentz invariant form factors
which are scalar functions of the momentum transfer $q^2 = (P-P^{'})^2$
where $P$ and $P^{'}$ are the four-momenta of the $B_c$, $B_s$ ($B_s^*$) mesons
respectively.

In our formalism, the mesons are taken as bound states of a quark and an
anti-quark. We construct the meson states as \cite{ISGW1}
\begin{eqnarray}
|M({\bf {P_M}},J,m_J)\rangle\  & = &
\sqrt{2M} \int d^{3}{\bf p} \langle L m_{L}S m_{S}|J m_J\rangle\  \langle
s m_s \bar{s} m_{\bar{s}}|S m_S\rangle\ \nonumber\\
 & &\Phi_{L m_L}({\bf p})|\bar q( {m_{\bar
q} \over \ M_{q \bar q} } {\bf {P}_M} - {\bf {p}},m_{\bar s})
\rangle|q({m_q \over \ M_{q \bar q} } {\bf {P}_M} + {\bf {p}},m_s)\rangle,
\label{states}
\end{eqnarray}
where the quark states are given by 
\begin{eqnarray}
|q({\bf {p}},m_s)\rangle\ &=& \sqrt{\frac{(E_q + m_q)}{2m_q}} \pmatrix{
 \chi^{m_s} \cr \ {{\bf{\sigma}}\cdot{\bf p }\over
{(E_q+m_q)}}  \chi^{m_s} \cr }, \nonumber\\
M_{q \bar q}&=&m_q+m_{\bar q}, \nonumber\\ 
E_q&=&\sqrt{m^2_{q}+{\bf{p}}^2}. 
\end{eqnarray}
 
In the above equations $M$ is the meson mass. The meson and the
constituent quark states satisfy the normalization conditions.
\begin{eqnarray} \langle
M({\bf{P^\prime}_M},J^{\prime},m^{\prime}_J)|M({\bf {P}_M},J,m_J)
\rangle\ &=& 2E\delta^3({\bf
{P^\prime}_M}-{\bf{P}_M})\delta_{J^{\prime},J}\delta_{m^{\prime}_J,m_J},
\end{eqnarray} \begin{eqnarray} \langle
q({\bf{p^\prime}},m^{\prime}_s)|q({\bf{p}},m_s) \rangle\ &=&
{E_q\over{m_q}} \delta^3({\bf {p^\prime}}-{\bf
{p}})\delta_{m^{\prime}_s,m_s}. \end{eqnarray}

The wavefunctions $\Phi_{L m_L}$ appearing in Eq.(\ref{states}) for the
mesons are calculated by solving reductions of Bethe-Salpeter equation
\cite {sommerer2}. We have applied this formalism to evaluate the
semileptonic form factors of the $B$ to $D$ and $D^*$ mesons and showed
that our results \cite{BD2} are consistent with the heavy quark
effective theory (HQET). We use wavefuctions from reduction B as
we did in our previous work on $B$ decays \cite{BD2}. 

The values of the Cabibbo-Kobayashi-Maskawa $(CKM)$ matrix elements we
use in this paper are $V_{ud}=0.974$, $V_{us}=0.2196$, $V_{ub}=0.0033$,
$V_{cd}=0.224$, $V_{cs}=0.974$, $V_{cb}=0.0395$ \cite{PDG}.

In Fig. 1 we show the semileptonic form factors for $B_c \rightarrow B_s
(B_s^{*})$ and in Table \ref{tp81} we show the exclusive semileptonic
decay widths to different pseudoscalar and vector final states ($B_c^+
\rightarrow P(V)e^+ \nu$). We also compare our results with those of
\cite{cc}. We notice in Fig. 1 that, although the semileptonic form factors are
qualitatively similar to the $B \rightarrow D (D^*) $ ones \cite{BD2},
flavor symmetry is absent in $B_c$ decays as discussed in \cite{jenkins}.

\begin{table}[h!tb]
\caption{
Exclusive semileptonic $B_c^+ \rightarrow P(V) e^+ \nu$ decay widths in $10^{-6}$ ev. 
}
\begin{center}
\begin{tabular}{|c|c|c|c|}
\hline\hline
& Process     & Decay width              & Decay width        \\
&             & This work                & Chang and Chen \cite{cc}         \\
\hline\hline
& $B_c^+ \rightarrow \eta_c e^+ \nu$                 &11.1 &14.2   \\
$\bar{b}$ decay& $B_c^+ \rightarrow J/\psi e^+ \nu $ &30.2 &34.4   \\
& $B_c^+ \rightarrow  D^0 e^+ \nu $                  &0.049&0.094  \\
& $B_c^+ \rightarrow  D^{*0} e^+ \nu  $              &0.192&0.269  \\
\hline\hline
& $B_c^+ \rightarrow  B^0_s e^+ \nu  $               &14.3 &26.6   \\
$c$ decay& $B_c^+ \rightarrow B_s^{*0} e^+ \nu  $    &50.4 &44.0   \\
& $B_c^+ \rightarrow B^0 e^+ \nu  $                  &1.14 &2.30   \\
& $B_c^+ \rightarrow B^{*0} e^+ \nu  $               &3.53 &3.32   \\
\hline\hline
\end{tabular}
\label{tp81}
\end{center}
\end{table}

For non-leptonic decays, the effective Hamiltonian (considering the
$B_c^+ \rightarrow B_s \pi^+$ channel) may be written as   
\begin{eqnarray}
H_{W} & = & \frac{G_F}{\sqrt{2}}V _{ cs } V^*_{ud} [ c_1(\mu) O_1 + c_2(\mu)
O_2 ],
\label{hnl} 
\end{eqnarray}
where
\begin{eqnarray}
O_1 & =& (\bar u_i d_i)_{V-A}(\bar s_j c_j)_{V-A},\nonumber\\
O_2 & =& (\bar u_i d_j)_{V-A}(\bar s_j c_i)_{V-A},
\end{eqnarray}
with $(i,j=1,2,3)$ denoting color indices and $V-A$ referring to
$\gamma_\mu(1-\gamma_5)$. $c_1(\mu)$ and $c_2(\mu)$ are short distance
Wilson coefficients computed at the scale $\mu$. By factorizing matrix
elements of the four-quark operator contained in the effective
Hamiltonian of Eq.(\ref{hnl}), one can distinguish three classes of decays
\cite{ns}. The first class ( class I ) contains those decays in which only
a charged meson can be generated directly from a color-singlet current,
as in $B_c^+ \rightarrow B_s \pi^+$. A second class of transitions ( class
II ) consists of those decays in which the meson generated
directly from the current is neutral, like the $\pi^0$ meson in the decay
$B_c^+ \rightarrow B^+ \pi^0$. Class I decay amplitudes are proportional to
$a_1$, class II decay amplitudes are proportional to $a_2$ where
\begin{eqnarray}
a_1 &=& c_1(\mu) + \xi c_2(\mu),\nonumber\\
a_2 &=& c_2(\mu) + \xi c_1(\mu),
\end{eqnarray}
and $\xi=1/N_c$ , where $N_c$ is the number of quark colors, and $\mu$ is
the scale at which factorization is assumed to be relevant. For the
third class ( class III ) the $a_1$ and $a_2$ amplitudes
interfere. Although the QCD factors $a_1$, and $a_2$ have been
calculated beyond the leading logarithmic approximation \cite{Buras}, we
will follow the prevailing convention of theoretical predictions and express
our results in terms of them. As an example the $B_c^+ \rightarrow B_s \pi^+$ amplitude takes the form
\begin{eqnarray}
A(B_c^+ \rightarrow B_s \pi^+) &=& \frac{G_F}{\sqrt{2}}V _{ cs } V^*_{ud}
 a_1(\mu) <\pi^+|(\bar u_i d_i)_{V-A}|0><B_s|(\bar s_j c_j)_{V-A}|B_c>.
\label{factor12}
\end{eqnarray}

The matrix elements $<B_s|(\bar s_j c_j)_{V-A}|B_c>$ in Eq.(\ref{factor12}) have already been
evaluated in semileptonic decays of the $B_c$ meson in terms of form factors, while the other matrix element
$(<\pi^+|(\bar u_i d_i)_{V-A}|0>)$ is related to
the decay constant of the relevant meson. The weak decay constants $f_P$
and $f_V$ for
pseudoscalar and vector mesons are defined by
\ber
<0|J_{\mu}|P(p)> & = & i f_P p_{\mu},\nonumber\\
<0|J_{\mu}|V(p)> & = & M_Vf_V \varepsilon_{\mu} , \
\eer
where $P$ and $V$ are pseudoscalar and vector states, respectively, and $J_{\mu} =  V_{\mu} -A_{\mu}$ is the weak current ($V_{\mu}$ and
$A_{\mu}$ are the vector and axial vector currents). The decay constants
can be expressed in terms of the wavefunctions of the relevant mesons and
are given by \cite{VI}
\ber
f_i & = & \sqrt{\frac{12}{M}}
\int^{\infty}_{0}\frac{p^2dp}{2 \pi^3}
\sqrt{\frac{(m_q + E_q)(m_{\bar{q}} + E_{\bar q})}{4E_qE_{\bar{q}}}}
F_i(p) , \\
F_{P}(p) & = & \left[ 1-\frac{p^2}{
(m_q + E_q)(m_{\bar{q}} + E_{\bar q})}\right]\psi_P(p), \\
F_V(p) & = & \left[ 1-\frac{p^2}{
3(m_q + E_q)(m_{\bar{q}} + E_{\bar q})}\right]\psi_V(p), \
\eer
where $\psi_{P(V)}$ are the momentum wavefunctions of the pseudoscalar (vector)
mesons.

We have previously applied this formalism to evaluate the decay
constants and the non-leptonic decays of the $B$ mesons \cite{BD3}. The
values of the decay constants we use in this paper are $f_\pi=0.130 $
GeV, $f_\rho=0.208 $ GeV, $f_K=0.159 $ GeV, $f_{K^*}=0.214 $ GeV,
$f_D=0.209 $ GeV, $f_{D^*}=0.237 $ GeV, $f_{D_s}=0.213 $ GeV,
$f_{D^*_s}=0.242 $ GeV, $f_{\eta_c}=0.400 $ GeV, $f_{J/\psi}=0.400 $
GeV. These values are the available experimental ones
\cite{PDG}. Otherwise we use our values reported in \cite{BD3}. These
values of the decay constants are similar to those used by other authors
\cite{GER, ex1, cc}.

In Table \ref{tp83} we compare our results for the exclusive non-leptonic $B_c \rightarrow PP, PV, VV$ decay
widths of different channels where the $\bar b$ quark
decays with those of \cite{cc}, while in Table \ref{tp84}, we make the
same comparison for the case of $c$ quark decays. At first glance, our
decay widths in Table \ref{tp83} are generally smaller than those of
Ref. \cite{cc} by 20-40$\%$. However, this is not a uniform trend as our
$B_c^+\rightarrow D^+\overline{D^{*0}}$ is 10$\%$ larger than that of
Ref. \cite{cc}. If we furthermore compare total lifetimes for $B_c$ we find that our
lifetime (0.46 ps) is longer compared to Ref. \cite{cc} (0.40 ps)
which is consistent with the dominant trends seen in the comparisons of the
exclusive channels. Both theoretical lifetimes are well within current
experimental uncertainties. Thus, experimental results for a set of
exclusive channels could resolve between these two sets of theoretical
predictions. Table \ref{tp84} displays even greater range of differences
between our model and that of Ref. \cite{cc}.

\begin{table}[h!tb]
\caption{
Exclusive non-leptonic decay widths of the $B_c$ meson in $10^{-6}$
ev. $\bar b$ quark decays with $c$ quark spectator. The authors of
Ref. \cite{cc} did not report the widths of some of the channels because
it was thought, prior to the experimental discovery of the $B_c$
meson, that these channels will be kinematically closed.}
\begin{center}
\begin{tabular}{|c|c|c|c|}
\hline\hline
Class& Process     & Decay width              & Decay width        \\
&             & This work                & Chang and Chen \cite{cc}         \\
\hline\hline
& $B_c^+ \rightarrow \eta_c \pi^+$      &$a_1^2 1.59$ &  $a_1^2 2.07 $   \\
& $B_c^+ \rightarrow \eta_c \rho^+$     &$a_1^2 3.74$ &  $a_1^2 5.48 $   \\
& $B_c^+ \rightarrow J/\psi \pi^+$      &$a_1^2 1.22$ &  $a_1^2 1.97 $   \\
& $B_c^+ \rightarrow J/\psi \rho^+$     &$a_1^2 3.48$ &  $a_1^2 5.95 $   \\
I & $B_c^+ \rightarrow \eta_c K^+ $     &$a_1^2 0.119$&  $a_1^2 0.161$   \\
& $B_c^+ \rightarrow \eta_c K^{*+}$     &$a_1^2 0.200$&  $a_1^2 0.286$   \\
& $B_c^+ \rightarrow J/\psi K^+$        &$a_1^2 0.090$&  $a_1^2 0.152$   \\
& $B_c \rightarrow J/\psi K^{*+}$       &$a_1^2 0.197$&  $a_1^2 0.324$   \\
\hline\hline
& $B_c^+ \rightarrow D^+ \overline{D^0}$           &$a_2^2 0.633$ & $a_2^2 0.664 $  \\
& $B_c^+ \rightarrow D^+ \overline{D^{*0}}$        &$a_2^2 0.762$ & $a_2^2 0.695 $  \\
& $B_c^+ \rightarrow  D^{*+} \overline{D^0}$       &$a_2^2 0.289$ & $a_2^2 0.653 $  \\
& $B_c^+ \rightarrow  D^{*+} \overline{D^{*0}}$    &$a_2^2 0.854$ & $a_2^2 1.080 $  \\
II & $B_c^+ \rightarrow D_s^+ \overline{D^0}$      &$a_2^2 0.0415$& $a_2^2 0.0340$  \\
& $B_c^+ \rightarrow D_s^+ \overline{D^{*0}}$      &$a_2^2 0.0495$& $a_2^2 0.0354$  \\
& $B_c^+ \rightarrow  D_s^{*+} \overline{D^0}$     &$a_2^2 0.0201$& $a_2^2 0.0334$  \\
& $B_c^+ \rightarrow  D_s^{*+} \overline{D^{*0}}$  &$a_2^2 0.0597$& $a_2^2 0.0564$  \\
\hline\hline
& $B_c^+ \rightarrow \eta_c D_s^+$             &$( a_12.16+ a_22.57)^2$&$( a_11.13+ a_21.98)^2$   \\
& $B_c^+ \rightarrow \eta_c D_s^{*+}$          &$( a_12.03+ a_22.16)^2$&$( a_11.04+ a_21.90)^2$   \\
& $B_c^+ \rightarrow J/\psi D_s^+$             &$( a_11.62+ a_21.72)^2$&$( a_11.02+ a_21.95)^2$   \\
& $B_c^+ \rightarrow J/\psi D_s^{*+}$          &$( a_13.13+ a_23.67)^2$&$                       $   \\
III& $B_c^+ \rightarrow \eta_c D^+$            &$( a_10.485+ a_20.528)^2$&$( a_10.193 + a_20.440 )^2$   \\
& $B_c^+ \rightarrow \eta_c D^{*+}$            &$( a_10.466+ a_20.452)^2$&$( a_10.181 + a_20.430 )^2$   \\
& $B_c^+ \rightarrow J/\psi D^+$               &$( a_10.372+ a_20.338)^2$&$( a_10.177 + a_20.442 )^2$   \\
& $B_c^+ \rightarrow J/\psi D^{*+}$            &$( a_10.686+ a_20.732)^2$  &$                   $   \\
\hline\hline
\end{tabular}
\label{tp83}
\end{center}
\end{table}
\begin{table}[h!tb]
\caption{Exclusive non-leptonic decay widths of the $B_c$ meson in $10^{-6}$ ev. $c$ quark
decays with $\bar b$ quark spectator.}
\begin{center}
\begin{tabular}{|c|c|c|c|}
\hline\hline
Class& Process     & Decay width              & Decay width        \\
&             & This work                & Chang and Chen \cite{cc}         \\
\hline\hline
& $B_c^+ \rightarrow B_s^0 \pi^+$              &$a_1^2 15.8$&$a_1^2 58.4$   \\
& $B_c^+ \rightarrow B_s^0 \rho^+$             &$a_1^2 39.2$&$a_1^2 44.8$   \\
& $B_c^+ \rightarrow B_s^{*0} \pi^+$           &$a_1^2 12.5$&$a_1^2 51.6$   \\
& $B_c^+ \rightarrow B_s^{*0} \rho^+$          &$a_1^2 171.$&$a_1^2 150.$   \\
I & $B_c^+ \rightarrow B_s^0 K^+$              &$a_1^2 1.70$&$a_1^2 4.20$   \\
& $B_c^+ \rightarrow B_s^{*0} K^+$             &$a_1^2 1.34$&$a_1^2 2.96$   \\
& $B_c^+ \rightarrow B_s^0 K^{*+}$             &$a_1^2 1.06$&$          $   \\
& $B_c^+ \rightarrow B_s^{*0} K^{*+}$          &$a_1^2 11.6$&$          $   \\
& $B_c^+ \rightarrow B^0 \pi^+$                &$a_1^2 1.03$&$a_1^2 3.30$   \\
& $B_c^+ \rightarrow B^0 \rho^+$               &$a_1^2 2.81$&$a_1^2 5.97$   \\
& $B_c^+ \rightarrow B^{*0} \pi^+$             &$a_1^2 0.77$&$a_1^2 2.90$   \\
& $B_c^+ \rightarrow B^{*0} \rho^+$            &$a_1^2 9.01$&$a_1^2 11.9$   \\
& $B_c^+ \rightarrow B^0 K^+$                  &$a_1^2 0.105$&$a_1^2 0.255$   \\
& $B_c^+ \rightarrow B^0 K^{*+}$               &$a_1^2 0.125$&$a_1^2 0.180$   \\
& $B_c^+ \rightarrow B^{*0} K^+$               &$a_1^2 0.064$&$a_1^2 0.195$   \\
& $B_c^+ \rightarrow B^{*0} K^{*+}$            &$a_1^2 0.665$&$a_1^2 0.374$   \\
\hline\hline
& $B_c^+ \rightarrow B^+ \overline{K^0}$       &$a_2^2 39.1$&$a_2^2 96.5$   \\
& $B_c^+ \rightarrow B^+ \overline{K^{*0}}$    &$a_2^2 46.8$&$a_2^2 68.2$   \\
& $B_c^+ \rightarrow B^{*+} \overline{K^0}$    &$a_2^2 24.0$&$a_2^2 73.3$   \\
& $B_c^+ \rightarrow B^{*+} \overline{K^{*0}}$ &$a_2^2 247$&$a_2^2 141$   \\
II & $B_c^+ \rightarrow B^+ \pi^0$             &$a_2^2 0.51$&$a_2^2 1.65$   \\
& $B_c^+ \rightarrow B^+ \rho^0$               &$a_2^2 1.40$&$a_2^2 2.98$   \\
& $B_c^+ \rightarrow B^{*+} \pi^0$             &$a_2^2 0.38$&$a_2^2 1.45$   \\
& $B_c^+ \rightarrow B^{*+} \rho^0$            &$a_2^2 4.50$&$a_2^2 5.96$   \\
\hline\hline
\end{tabular}
\label{tp84}
\end{center}
\end{table}

In conclusion, we have systematically evaluated the decay widths of the
exclusive semileptonic channels $B_c \rightarrow P(V) e \nu$ and the
exclusive two-body non-leptonic decays $B_c \rightarrow PP$, $PV$, and
$VV$ assuming that either $c$ or $\bar b$ quark inside the $B_c$ meson
is a spectator quark, and using our relativistic model
\cite{sommerer2}. In general, our predicted widths are smaller than those
reported in Ref. \cite{cc} but there are exceptions to this trend. The
variations between the theoretical predictions are wide enough so that
experimental results should be able to discern between the models.

We note that the dominant decays are those when the $b$ quark inside the
$B_c$ meson behaves as a spectator quark and a vector meson is produced
in the final state. In fact, $B_c^+ \rightarrow B_s^{*0}e^+\nu$ is the
dominant decay among all the semileptonic channels ( see Table 2 ) and
$B_c \rightarrow B_s^{*0}\rho^+$ becomes the dominant among all the
two-body non-leptonic decays ( see Table 4 ). Although these decays are
suppressed by phase space they are CKM favored.

Finally we point out that the ratio
\begin{equation}
\frac{\Gamma(B_c^+ \rightarrow V\rho^+)}{\left. d\Gamma(B_c^+
\rightarrow Ve^+\nu)/dt \right|_{t=m_{\rho}^2}}=6 \pi^2 | V_{ud} |^2
a_1^2 f^2_{\rho},
\end{equation}
with $V= B_s^{*0}, \ J/\psi$ will be a good experimental test for the
numerical value of the coefficient $a_1$ of QCD \cite{ns}.

{\bf Acknowledgments}

This work was supported in part by the US Department of Energy, Grant
No. DE-FG02-87ER40371, Division of High Energy and Nuclear Physics and
by the International Institute of Theoretical and Applied Physics, Ames,
Iowa. The work of A.A. was also supported in part by the DOE under
contract number DE-FG02-92ER40730.

{\bf Figure Caption}

\begin{itemize}
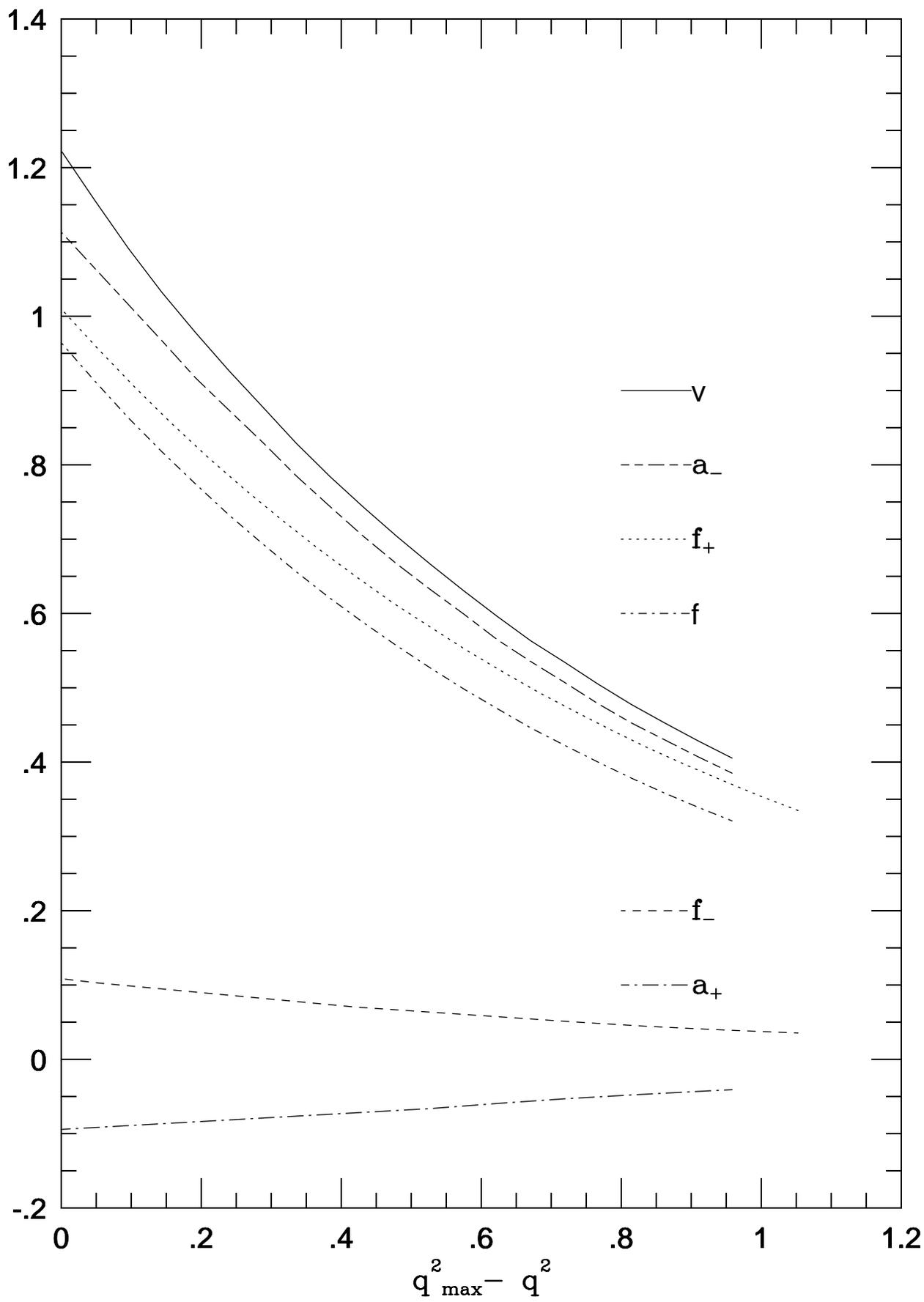
\item[\bf{Fig. 1}]

The semileptonic form factors for $B_c \rightarrow B_s (B_s^{*})$.

\end{itemize}


\begin{thebibliography}{References}


\bibitem{CDF}
CDF Collaboration, F. Abe {\it et al.}, Phys. Rev. D {\bf 58}, 112004 (1998); Phys. Rev. Lett. {\bf 81}, 2432 (1998). 


\bibitem{review}
S.S.~Gershtein, {\it et al.},
hep-ph/9803433.

\bibitem{pp}
A. Abd El-Hady, M.A.K. Lodhi, and  J.P. Vary, 
Phys. Rev. D {\bf 59}, 094001 (1999). 


\bibitem{sommerer2}
A. J. Sommerer, J. R. Spence, and J. P. Vary, Phys. Rev. C {\bf 49}, 513
(1994); A. J. Sommerer, A. Abd El-Hady, J. R. Spence, and J. P. Vary, 
Phys. Lett. B {\bf 348}, 277 (1995). 



\bibitem{itzykson}C. Itzykson, and J.B. Zuber, Quantum Field theory.
McGraw-Hill, New York (1980) (Chapter 10 gives a review of
Bethe-Salpeter equation); G. E. Brown and A. D. Jackson, The Nucleon-Nucleon
Interaction North-Holland, New York (1976); and references therein; B. Silvestre-Brac, A. Bilal, C. Gignoux, and P.
Schuck, Phys. Rev. D {\bf 29}, 2275 (1984).



\bibitem{Thompson}R. H. Thompson, Phys. Rev. D {\bf 1}, 110 (1970).

\bibitem{sommerer}J. R. Spence and J. P. Vary, Phys. Rev. C {\bf  47}, 1282
(1993); A. J. Sommerer, J. R. Spence, and J. P. Vary, Mod.
Phys. Lett. A {\bf  8}, 3537 (1993).




\bibitem{EQ}
E. J. Eichten and C. Quigg, Phys. Rev. D {\bf 49}, 5845 (1994). 


\bibitem{GER}
S.S. Gershtein, {\it et al.}, hep-ph/9803433; S. S. Gershtein, V.V. Kiselev,
A.K. Likhoded, and A.V. Tkabladze, Phys. Rev. D {\bf 51}, 3613 (1995); Phys. Usp. {\bf 38}, 1 (1995).


\bibitem{ex1}
D.~Du and Z.~Wang,
Phys.\ Rev.\ D {\bf 39}, 1342 (1989);
 M.~Lusignoli and M.~Masetti,
Z.\ Phys.\ C {\bf 51}, 549 (1991);
 A. Y. Anisimov, P. Y. Kulikov, I. M. Narodetskii, and K. A. Ter-Martirosyan, hep-ph/9809249 (accepted for publication in Phys. Atom. Nucl.);
 D. Du, G. Lu and Y. Yang, Phys. Lett. {\bf B387}, 187 (1996).

\bibitem{cc}
C-H.  Chang and Y-Q. Chen,
Phys. Rev. D {\bf 49}, 3399 (1994). 



\bibitem{ISGW1}
N.  Isgur, D.  Scora, B. Grinstein, and M. B. Wise, Phys. Rev. D 
{\bf  39}, 799 (1989).


\bibitem{BD2}
A. Abd El-Hady, A. Datta, K.S. Gupta, and  J.P. Vary, 
Phys. Rev. D {\bf 55}, 6780 (1997). 

\bibitem{PDG}
Particle Data Group, C.~Caso {\it et al.},
Eur.\ Phys.\ J.\ C {\bf 3}, 1 (1998).


\bibitem{jenkins}
E. Jenkins, M. Luke, A.V. Manohar, and M.J. Savage,
Nucl. Phys.  {\bf B390}, 463 (1993)




\bibitem{ns} 
M. Neubert and B. Stech, in {\it Heavy Flavors}, 2nd ed., edited by
A. J. Buras and M. Lindner (World Scientific, Singapore, 1998), hep-ph/9705292.

\bibitem{Buras}
A.J.~Buras,
Nucl.\ Phys.\ {\bf B434}, 606 (1995)

\bibitem{VI} S. Veseli and I. Dunietz, Phys. Rev. D {\bf  54} 6803 (1996).


\bibitem{BD3}
A. Abd El-Hady, A. Datta, and  J. P. Vary, 
Phys. Rev. {\bf D58}, 6780 (1998). 



\end{thebibliography}
\end{document}